# Noninvasive Extraction of Maternal and Fetal Electrocardiograms Using Progressive Periodic Source Peel-off

Yao Li#, Xuanyu Luo#, Haowen Zhao, Jiawen Cui, Yangfan She, Dongfang Li,
Lai Jiang*, Xu Zhang*, *Member*, *IEEE*

*Abstract*—Abdominal electrocardiogram (AECG) gives a safe and non-invasive way to monitor fetal well-being during pregnancy using surface electrodes. However, it is challenging to extract weak fetal ECG (fECG) from the AECG recordings with larger maternal ECG (mECG) and external noises. In this study, we introduce a novel progressive periodic source peel-off (PPSP) method for extracting periodic ECG sources from multi-channel AECG recordings, including three main modules: 1) A periodic constrained FastICA (PCFICA) module with ECG physiology-informed constraints for extracting precise ECG spike trains, 2) A singular value decomposition module for estimating ECG waveforms, and 3) A peel-off strategy that facilitates to discern weak fECG source by eliminating previously separated sources or noises. The performance of the PPSP method was examined on two public databases, synthetic data and our clinical data. For extracting fECG spike trains, our PPSP method achieved an F1-scores of 99.59% on public data, 99.50% on synthetic data at the highest noise level. It further yielded the lowest RMSE of fetal heart rate of 6.20% on clinical data. It significantly outperformed other state-of-the-art methods on any set of data ($p < 0.05$). This study demonstrated effectiveness of the PPSP method for extracting and separating mECG and weak fECG signals, with high precision especially at high noise levels. Our study promotes noninvasive measurement and intelligent monitoring of both fetal and maternal heart activities towards advanced healthcare in perinatal medicine.

## I. Introduction

ANNUALLY, around 2 million stillbirths take place worldwide [1]. The monitoring of fetal heart activities plays a crucial role in reducing perinatal morbidity and mortality associated with a wide range of fetal heart anomalies [2]. To obtain clinical information on the fetal heart activities non-invasively, many electronic devices have been applied for fetal cardiac assessment, including cardiotocography (CTG) [3], Doppler ultrasound [4], fetal magnetocardiography (MCG) [5], fetal phonocardiography (PCG) [6] and fetal electrocardiography (ECG), also termed fECG [7]. Although the CTG has been widely used, it is susceptible to signal loss and inaccurate heart rate estimation [8]. Both the CTG and the Doppler ultrasound carry inherent safety concerns due to the use of high-frequency ultrasound signals directed towards the fetus [9, 10]. The fetal PCG, while useful, is sensitive to sensor placement and requires extensive data processing to distinguish fetal cardiac signals from background noises [11]. Considering the limitations of existing tools, the fECG emerges as a safe and recommended technology for sensing fetal cardiac activities [12], offering precise beat-to-beat fetal heart rate (FHR) acquisition and promising prospects for long-term monitoring.

Although the fECG technology is relatively safe, convenient and capable of reflecting fetal heartbeat in a noninvasive manner, it still faces challenges. As fECG signals propagate from the body to the electrodes placed over maternal abdominal skin surface, they pass through a series of body tissues, with greatly attenuated amplitudes [13]. Moreover, the actually recorded abdominal electrocardiogram (AECG) signal is always mixed with various noises, including maternal ECG (mECG), fetal peristalsis, electromyography (EMG) from belly muscles, power line interference and some motion artifacts, overlapping the fECG in both frequency and time domains [14-16]. Therefore, it is crucial to extract the R-peak from such contaminated signals for evolving the fECG technology towards fetal cardiac monitoring.

In order to make convenient, non-invasive fetal heart monitoring available via surface-collected fECG signals, researchers have explored numerous methods for extracting fECG signals precisely from noisy environments [17-19]. Adaptive filtering (AF) methods [20] were used to suppress the influence of mECG with the help of a reference signal recorded over the chest. However, literature [21, 22] reports that the final fECG still contained more or less mECG signal after the above filtering process. Template subtraction (TS) serves as a simple but insufficient idea to eliminate mECG components by subtracting a preset mECG template from the original signals. However, the average mECG template may be distorted by overlapping or incorrectly detected heartbeat cycles, and the mECG morphology is also sensitive to many factors including electrode configuration and body posture [19]. Some wavelet-based methods [23, 24] were performed, by making use of intrinsic multifaceted property to manipulate R-peak identification. Nevertheless, a suitable wavelet base

This work was supported in part by the Anhui Provincial Key Research and Development Project under Grant 2022k07020002. (*Corresponding authors: Lai Jiang and Xu Zhang.*)

# Both authors contribute equally to this work.

Y. Li, H. Zhao, Y. She, D. Li, X. Zhang are with the School of Microelectronics at University of Science and Technology of China, Hefei, Anhui 230027, China. (e-mail: xuzhang90@ustc.edu.cn).

X. Luo, J. Cui are with the School of Information Science and Technology at University of Science and Technology of China, Hefei, Anhui 230027, China.

L. Jiang is with the Department of Obstetrics and Gynecology at the First Affiliated Hospital of University of Science and Technology of China (Anhui Provincial Hospital), also with Division of Life Sciences and Medicine at University of Science and Technology of China, Hefei, Anhui 230001, China. (e-mail: gyobjiang@163.com)

has to be selected in prior, leading to low robustness. Deep learning (DL) methods [12, 25, 26] have also been investigated for extracting nonlinear features characterizing both fECG and mECG from AECG simultaneously. Although good performance has been reported with the DL solution, it is a very time-consuming approach heavily relying on a large number of training data.

Due to the nature of the collected electrophysiological signals being a superposition of different sources, we can also consider using blind source separation (BSS) methods for extracting fECG signals. Independent component analysis (ICA) [18], principal component analysis (PCA) and non-negative matrix factorization (NMF) [27] have been investigated to complete the task of fECG extraction. Although traditional BSS methods operated well in some of records, they may occasionally fail to detect a clean fECG sufficiently [28]. By incorporating a reasonable assumption of pseudo-periodicity, periodic component analysis ($\pi$CA) has been developed with significantly improved performance [29]. However, it still struggles to perform well in noisy environments due to lack of sufficient understanding and precise description of the concerned signal sources.

In recent years, a novel BSS method known as progressive FastICA peel-off (PFP) was introduced within the realm of surface electromyogram (SEMG) decomposition [30-33]. This method have addressed a range of challenges in decomposing SEMG into its constituent sources termed motor unit (MU) action potential trains by employing an innovative framework. In the PFP framework, the use of a constrained FastICA is a key step of refining more accurate MU spikes from the spikes initially extracted by the common FastICA approach. Furthermore, a peel-off strategy is utilized to avoid local convergence to larger MU spikes, thus facilitating to extract more weak MU activities. Inspired by their work, we can learn from their ideas to tackle the challenges we face in extracting weak fECG signals.

This study presents a modified version of the PFP method specifically designed for extracting periodic ECG-like electrophysiological sources, termed progressive periodic source peel-off (PPSP). In this PPSP method, the progressive peel-off strategy was adopted to sufficiently mine weak fECG source signal components under strong mECG interferences. In addition, a novel periodic constrained FastICA (PCFICA) algorithm was developed to detect and correct the mECG and fECG spikes during the primary source separation process. Our method gives a path to non-invasive and reliable detection and monitoring of both fECG and mECG signals in clinical practice, plays a crucial role in ensuring the health of the fetus and pregnant women, as well as promoting intelligent clinical monitoring in perinatal medicine.

## II. METHODS

### A. Dataset Description

I）. Public databases

Two public databases were utilized for method validation in this study, as described below.

1) ADFECG. ADFECG database [34] has 5 multichannel records, acquired from 5 subjects between 38 and 41 weeks of gestation. Each record composed 4-channel AECG signals taken from pregnant women's abdomen, as well as one channel of reference signal considered as ground-truth fECG signal directly taken from fetal scalp, all signals are sampled at 1 kHz. Annotations of fetal R-peak timings were verified by medical experts and provided in the dataset.

2) NIFECGA. There are 75 AECG records in Set-A of the 2013 Physionet/Computing in Cardiology Challenge [35]. Each record has four-channel AECG signals and one observation/reference channel fetal R-peak annotations obtained from direct fetal scalp ECG by proficient cardiologists. As suggested by previous works [10, 11, 25], a few records namely a33, a38, a52, a54, a71, and a74 were discarded from the database due to inaccurate fetal R-peak annotations, leaving 69 records for assessment. Each recording is with a duration of 60 s and a sampling rate of 1 kHz.

II）. Clinical data

All clinical data used in this study were collected at Department of Obstetrics in the First Affiliated Hospital of University of Science and Technology of China (Hefei, Anhui, China). All protocols of this study were approved by the Ethics Review Committee for Medical Research in the hospital, under Application No. 2024KY-188. Five women with healthy singleton pregnancy at gestational ages of 36-40 weeks participated into the data collection experiments. Informed and signed consent was obtained from all participants or their conservators before any procedure of the experiments.

During the experiments, all pregnant participants were lay down on an examination bed quietly for AECG recordings. A flexible 4×4 grid electrode array was placed over the upper abdomen of each participant, as shown in Figure 1. Each electrode has a round contact in a diameter of 8 mm, enabling one channel of data recording. The spacing between electrodes is 17.5 mm. For each pregnant participant, the experiment consisted of 5 sessions, with each session lasting for 15-20 minutes. During the data recording, each participant was asked to just keep relaxed. Some slight movements or body posture adjustments were allowed.

The electrode array was connected via a tiny cable to a light device functioning as the data recording system (FlexMatrix Inc., Shanghai, China). It was built with a two-stage amplifier at a total gain of 24 dB, a band-pass filter set at 1-100Hz for each channel and an analog-to-digital converter (ADS1299, Texas Instruments, Texas, USA), at a sampling rate of 200Hz. The system also enabled wireless data transmission via a Bluetooth connection to a model device like cell phone or pad. A software application with graphical user interface was also developed for real-time signal monitoring and storage. All data were divided into segments of 1-3 minutes in length for subsequent analyses.

For all datasets, before processed by the proposed PPSP methods, several preprocessing steps were undertaken for simple noise reduction. The impulsive artifacts were removed from each AECG channel, as suggested in [36]. After that, a digital band-pass filter at 3-100 Hz was further applied to each AECG channel. Power-line interference as well as its harmonics were also removed using a series of digital notch filters.

III）. Synthetic data

Synthetic AECG recordings were also generated where clean fECG and pure mECG signals were known a prior. The fECGSYN simulator described in [37], [38] was used due to its convenience and good simulation of fetal-maternal mixtures. The fECG signal, mECG signal and other noise were considered as three individual signal sources derived from the Fetal ECG Synthetic Database (fECGSYNDB) [37]. These source signals were assumed to propagate onto the observational sites (skin-electrode contacts) to synthesize one channel of the AECG signal. We intentionally simulate 16 AECG channels in a 4×4 matrix over the abdominal region, which was consistent with the array used in the clinical data collection experiments. The signal-to-noise ratio (SNR) between fECG and mECG signals was considered to be relatively constant in most practical cases, we chose the SNR of the fECG relative to mECG ($SNR_{fm}$) to be a constant and reasonable value of -9 dB. On this basis, five sets of AECG data were generated when the SNR of the mECG over noise ($SNR_{mn}$) was set to 6, 3, 0, -3 and -6 dB, corresponding to the SNR of fECG over all noises ($SNR_{fn}$) at -3, -6, -9, -12, -15 dB, respectively. For more details, please refer to [38]. For each set, there were 8 recordings (representing the data from 8 different subjects) sampled at 250 Hz, each lasting 3 minutes for subsequent data analyses.

*B. fECG Extraction*

This study presents a modified version of the PFP method specifically designed for extracting periodic ECG-like electrophysiological sources, termed PPSP. In this PPSP method, the progressive peel-off strategy was adopted to sufficiently mine weak fECG source signal components under strong mECG interferences. In addition, a novel PCFICA algorithm was developed to detect and correct the mECG and fECG spikes during the primary source separation process. Figure 2 shows the block diagram of our proposed PPSP method for weak fECG extraction, following the general PFP framework including three primary parts: (a) two-step (common and constrained) FastICA for source separation, (b) source signal waveform estimation and construction, and (c)

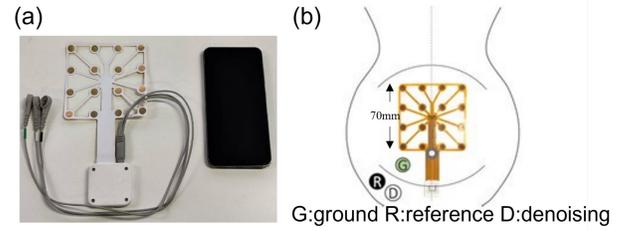

Figure 1 (a) Data recording system. (b) Illustration of electrode placement.

progressive peel-off of obtained source signals for iterative extraction of weaker sources. More details regarding the original PFP framework can be found in [30-33]. Substantial modifications and updates that boost its evolution into the PPSP method for fECG extraction are described as follows.

*1) Two-step FastICA for Precise Spike Detection*

The AECG signal is collected in a form of $m$ channels, preprocessed with band-pass and band-stop filtering and data segmentation (detailed in above dataset description), and denoted as $X(t) = [x_1(t), x_2(t), ..., x_m(t)]^T$ conform to the convolutive model [39]. The processed AECG signal was initialized as the residual signal, and then fed into subsequent BSS operations primarily relying on the two-step FastICA. A common FastICA was applied to the extended and whitened input signal to get preliminary source signals which is describe as [40]:

$$w^+ = E\{xG'(w^T x)\} - E\{G''(w^T x)\}w$$
$$w = w^+ / \|w^+\|_2 \tag{1}$$

until $|w_{k+1} - w_k| < \theta$, where $\theta$ is the convergence threshold, $w$ is the unmixing matrix, $x$ is the input signal, $G$ is a nonquadratic function where we can use $G(x) = \log(\cosh(x))$ as default. The source signals were extracted by iterative threshold Otsu algorithm [41] and sampling points with amplitude lower than the threshold were labeled as no R-peak firing event. From these preliminary source signals, one or some ECG-like source signals (either mECG or fECG or their constituent components) were determined and selected through the clustering and selection module. Then the initial spike trains were clustered by valley seeking approach to distinguish the spikes from the same source signal.

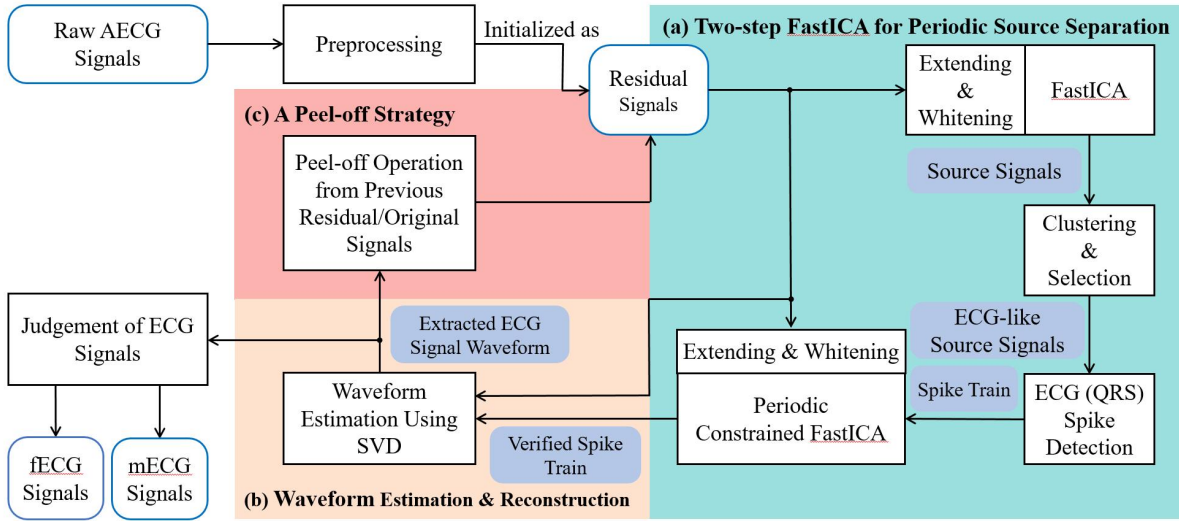

Figure 2. Flowchart of proposed method for fECG and MECG extraction based on general PFP framework consist of: (a) two-step FastICA for precise spike detection, (b) ECG source signal waveform estimation, and (c) a peel-off strategy. The red boxes indicating our innovative designs specific for MECG and fECG extraction.

The output of FastICA was regarded as a set of preliminarily separated source signals along with various noise contaminations, which did not correspond to the accurate physiological sources of interest. From these source signals, a rough but insufficiently accurate ECG spike sequence could be detected. It was essential to make efforts towards sufficiently accurate ECG spike train so as to obtaining its precise waveforms, based on the physiological characteristics of the ECG signals. Therefore, we introduced a new PCFICA process, using the rough ECG spikes as a reference input to obtain an accurate and reliable ECG spike train from the original multi-channel signals. Specifically, both the equality constraint and periodicity constraint were considered in the PCFICA process. The equality constraint described the use of initial spike train to guarantee convergence towards its corresponding ECG source. The periodicity constraint represented a loose constraint using autocorrelation that allowed the algorithm to converge towards the source with sufficient periodic activities. With both constraints mentioned above, the general problem of PCFICA is given as:

$$\begin{aligned} \max \quad & J_G(w) = [E\{G(w^T x)\} - E\{G(v)\}]^2 \\ s.t \quad & g_1(y) = \xi_1 - E\{y^T r\} \leq 0 \\ & g_2(y) = \xi_2 - E\{y^k y\} \leq 0 \\ & H(w) = \|w\|^2 - 1 = 0 \end{aligned} \quad (2)$$

where $J_G(w)$ is the same objective function as defined in Eq. 1; $g_1(y)$ is an equality constraint that measures the correlation between the estimated independent component $y = w^T x$ and the reference spike train $r$; $g_2(y)$ denotes the periodicity constraint with respect to the signal $y$ and its delay signal $y^k$ for

---

Algorithm 1. The PPSP method for fECG extraction
---
1: Initialize residual signal $\bar{\bar{x}}$ as AECG. Initialize candidate spike train set $\varphi = \emptyset$. Flag=1.
2: Extend and whiten $\bar{\bar{x}}$ then perform FastICA to extract spike train $\tilde{v}_1, \tilde{v}_2, ..., \tilde{v}_n$
3: **while** Flag=1 **do**
 Flag=0
 **for** i=1;i<N+1;i++ **do**
  Apply PCFICA on $\bar{\bar{x}}$ using $\tilde{v}_i$ as constraint to obtain calibration version (termed as $\tilde{\tilde{v}}_i$) of $\tilde{v}$
  **if** $Fr(\tilde{\tilde{v}}_i)$<1.11Hz or $Fr(\tilde{\tilde{v}}_i)$>3.33Hz
   continue
  end if
  **while** $\varphi_j$!=None:
   **if** COR($\tilde{\tilde{v}}_i, \varphi_j$)>0.5
    continue
   end if
   **if** IOQ($\tilde{\tilde{v}}_i$)>6 or SC($\tilde{\tilde{v}}_i$)<5
   Flag=1
    add $\tilde{\tilde{v}}_i$ into $\varphi_j$
    estimate the waveform $\vartheta$ of $\tilde{\tilde{v}}_i$
    update $\bar{\bar{x}} = \bar{\bar{x}} - \vartheta$ and **continue**
   end if
 end for
 Extend and whiten $\bar{\bar{x}}$ then perform FastICA to extract spike train $\tilde{v}_1, \tilde{v}_2, ..., \tilde{v}_n$
4: Select maternal and fetal ECG spike train from $\varphi$.

---

Table 1 The determination of modules in the ablation study

| Method | equality constraint | periodicity constraint | SVD |
|---|---|---|---|
| cfICA | √ | - | - |
| pcfICA | √ | √ | - |
| Proposed PPSP | √ | √ | √ |

time lag $k = 0,1, ..., N - 1$; $\xi_j (j = 1,2)$ denotes a threshold representing the lower bound of each constraint. The optimization problem was solved by augmented Lagrangian method. After the PCFICA converged, the calibrated spike train was obtained using R-peak detection procedure proposed in [42].

*II) Waveform Estimation Using SVD*

From the reliable ECG spike train obtained from the above process, SVD was then used to reconstruct corresponding ECG waveform [42]. For each channel in signal with a $q$-beat period, we calculated the mean sample points length of RR interval $p$ and the number of QRSs $q$ to construct a $p \times q$ ($p > q$) matrix $Z$. Then $Z$ was decomposed by SVD to obtain the singular value matrix and two orthogonal matrices:

$$Z = USV^T \quad (3)$$

where $S = [diag(\delta_1, \delta_2, ..., \delta_q, 0)]$, $\delta_1 > \delta_2 > ... > \delta_q$ denotes singular value matrix; $U = [u_1, u_2, ..., u_q]$, $v = [v, v_2, ..., v_q]$ denotes the left and right singular matrix. $Z$ can be indicated by the multiplication of $u_i$, $v_i$ and $\delta_i$, while $\delta_i$ represents the energy distribution of different matrix. Then the matrix $Z_r$ of approximated QRS waveform was rebuilt using a reduced number of eigenvectors:

$$Z_r = U_r S_r V_r^T \quad (4)$$

where the matrix $S_r$ of dimension $ne \times ne$ is the diagonal matrix of the first '$ne$' singular values, and $U_r(nd \times ne)$ and $V_r(nq \times ne)$ are the matrices of the first '$ne$' left and right singular vectors, respectively. $ne$ is the subspace of dimension. Empirically, $ne = 3$ was selected. For each channel of the residual signal, we use SVD to estimate the corresponding ECG signal waveform separately.

*III) Peel-off Strategy for Weak fECG Separation*

By using the two-step combination of FastICA and PCFICA, one or more desired sources were expected to be separated from the signals. However, considering that FastICA is easily affected by initialization, it usually converges to local optimal value. Therefore, it was difficult for FastICA to find weak fECG during a single round of source separation. To solve this problem, a feasible approach is to remove the extracted source signal and use the remaining signal for separation in order to uncover potential weak sources, which is known as the peel-off strategy [33]. It can be described as following in this study:

$$X_i^{(next)} = X_i^{(current)} - \tilde{X}_i \quad (5)$$

where $X_i^{(next)}$ denotes the updated residual signal, $X_i^{(current)}$ denotes the residual signal in current iteration and $\tilde{X}_i$ denotes the estimated waveform of mECG or fECG in this iteration using SVD, all of $i$ th channel. The peel-off strategy was introduced for each channel to mitigate the influence of identified ECG spike trains.

*IV) ECG Spike Train Reliability Judgement*

In the iterative separation process described above, each round yields the separated source (i.e., the peeled-off part $\widetilde{X}_i$). It was essential to determine whether the estimated signal was the desired ECG source. This judgment process includes assessing whether it is a genuine ECG signal and identifying whether it is mECG or fECG.

The following criteria were employed: (1) If the R-peak firing rate (denote as *Fr*) of the estimated spike train was lower than 1.11Hz or higher than 3.33Hz, this source signal was abandoned to avoid potential interference. (2) If the correlation (denoted as COR) between two spike trains higher than 0.5, we processed the duplicated spike train by discarding one of them. (3) If the spike train with the index of quality (denoted as IOQ) for the best mECG component identification higher than 6, we added the spike train into φ. (4) The stability coefficient (denoted as SC) which composed of the variation of R-R interval (denoted as $cov_{rri}$) and the variation of ECG spike amplitude (denoted as $cov_{amp}$) was defined as follows: $SC = \alpha_1 cov_{rri} + \alpha_2 cov_{amp}$, where $\alpha_1$, $\alpha_2$ were set to 100,1 respectively. If SC was lower than 5, we added the spike train into φ.

For the above criteria, once a validity spike train was added into φ, the residual AECG signals were update by employing peel-off procedure to subtract the waveform of the identified spike train from itself. The pseudocode of the proposed fECG extraction framework is illustrated in Algorithm 1.

C. *Performance Evaluation*

For public dataset and synthetic data, the pure fECG signal or the ground-truth fECG spike train (from direct fetal scalp ECG) are available. The following metrics were used to evaluate the performance of extracting the fECG or mECG: positive predictive value (PPV), sensitivity (Sen), accuracy (ACC) and F1 score, which are defined as:

$$Sen = \frac{TP}{TP + FN} \times 100\%$$

$$PPV = \frac{TP}{TP + FP} \times 100\% \quad (6)$$

$$ACC = \frac{TP}{TP + FP + FN} \times 100\%$$

$$F1 = \frac{2 \times PPV \times Sen}{PPV + Sen} \times 100\%$$

where TP, FP, and FN denote true positives, false positives, and false negatives, respectively. TP denote the number of correctly detected R-peak, FN denote the number of missed R-peak and FP denote the number of incorrect detected R-peak. According to [37], if the extracted ECG spike is within 50 ms from the fetal R-peak annotation, it is counted as a fetal R-peak prediction.

For the synthetic signals, due to the ground truth fECG waveform is available, we also use SNR to evaluate the level of noise suppression after applying different methods on synthetic data with different noise levels, which is defined as:

$$SNR = 10\log \frac{\|FECG\|^2}{\|\widehat{FECG} - FECG\|^2} \quad (7)$$

For synthetic data, fECG root mean square error (RMSE$_{fECG}$) is also used to assess the degree of difference between the waveform of extracted fECG and the ground truth. RMSE$_{fECG}$ is defined as:

$$RMSE_{FECG} = \sqrt{\frac{\sum_{i=1}^{n}\left[\widehat{FECG_i} - FECG_i\right]^2}{n}} \times 100\% \quad (8)$$

where $FECG_i$ represents the $i$ th point of ground truth fECG waveform, $\widehat{FECG_i}$ represents the $i$ th point of estimated fECG waveform, and $n$ represents the total sample number. The smaller the RMSE value, the closer the estimated fECG is to the reference, demonstrating the method's greater accuracy.

Three representative comparison methods were selected besides proposed method, which includes FastICA [42], representing traditional blind source separation methods, LMS [43], representing adaptive filtering, and CycleGAN [25], a commonly used deep learning method.

In addition, we used ablation experiments on public datasets to determine the necessity of each component in the proposed fECG extraction framework. As shown in Table 1, the ablation experiments included two different constraint modules and two different waveform estimation modules. Firstly, for each method, FastICA was used to output preliminary separations of mECG and fECG spike trains. Based on this, we first devised the PFP method, with the constrained FastICA (cfICA) module, which employed equality constraints for reliability assessment, while the waveform estimation method remained the least squares method originally used in PFP to estimate the averaged waveform of spike locations. The second method, with the PCFICA module, utilized periodicity constraints for reliability assessment, while keeping other parts unchanged. For all methods, the final fECG extraction performance was evaluated. Since PCFICA method and the proposed PPSP method differ only in waveform reconstruction method, we evaluated the impact of the reconstruction methods on the algorithms using synthetic signals by calculating the SNR and RMSE of the waveforms before and after reconstruction, using PCFICA and PPSP method.

D. *Statistical Analysis*

In order to better examine the performance of different fECG extraction methods (six levels: FastICA, LMS, CycleGAN, PFP, PCFICA and the proposed PPSP method) when processing public data and synthetic data, a series of one-way repeated-measure ANOVAs were applied on PPV, Sen, ACC, F1 score and RMSE$_{fECG}$, respectively. If necessary, multiple pairwise comparisons with LSD corrections were performed. The level of significant difference was set as $p < 0.05$. All statistical analyses were performed using SPSS software (version 27.0, SPSS Inc. Chicago, IL, USA) in this study.

III. RESULTS

A. *Results of Public Databases*

Figure 3 illustrates an example of extracting both mECG and fECG signals from a 15-s segment of multi-channel AECG signals showing two typical channels (AECG3-4) from the ADfECG database, using the proposed PPSP method. From visual inspection, both mECG and fECG signals can be separated progressively from original noisy AECG recordings. Specifically, the preliminary separated fECG source signal (Fig. 3d) is not sufficiently precise because some missing or erroneous spikes are carried. These errors presented in the fECG spike train are corrected after applying the PCFICA approach, leading to correct fECG signal and its precisely detected spikes presented in Fig. 3 (e). It can be observed that miss or incorrect spikes in the preliminary fECG are corrected, resulting in a more accurate spike train.

Figure 4 shows four quantitative metrics for evaluating both mECG and fECG extraction performance using six different methods. The results derived from public databases are listed in Fig. 4a. The proposed method achieved a SEN of (99.66 ± 0.37)%, a PPV of (99.53 ± 0.10)%, an ACC of (99.19 ± 0.44)% and an F1 score of (99.59 ± 0.22)%, averaged over all data segments in the public databases. The ANOVA also reports that our method outperformed any other comparative method with statistical significance ($p < 0.05$).

*B. Results of Synthetic Data*

Figure 5 shows examples from synthetic data with different $SNR_{mn}$ levels when the proposed method is applied for mECG and fECG extraction. It can be observed that the extracted mECG and fECG are consistent with their previously known sources (pure mECG and fECG sources in Figures 5 (a) and 5 (b)), respectively, regardless of any $SNR_{mn}$ level. Specifically, as the signal-to-noise ratio worsens from 6 dB to -6 dB, it becomes difficult to observe any clear ECG spikes from the mixed AECG signals. However, our method performed well even at the lowest $SNR_{mn}$ level of -6 dB, effectively extracting clear mECG and fECG waveforms.

The performance evaluation of all methods for extracting fECG from the synthetic data can be seen in Figure 4(b)-(f) where the pure fECG is used as ground truth. At relatively higher $SNR_{mn}$ levels, all methods demonstrated satisfactory results, with the proposed method showing a slight advantage without significant difference. However, as the $SNR_{mn}$ further decreases, the performance of the morphology-based LMS methods deteriorates significantly. In contrast, the FastICA method shows relatively less performance decline. The classic PFP and PCFICA methods perform better than FastICA, while the deep learning-based CycleGAN method is comparable to them. However, it still cannot achieve good extraction results when the entire data segment is covered by significant noise and artifacts, thereby affecting the overall extraction performance. Overall, the proposed method, which integrates PCFICA and SVD, achieved metrics that surpassed all other methods with statistical significance ($p<0.05$, across all $SNR_{mn}$ levels except 6dB), especially showing no significant performance degradation under low $SNR_{mn}$ conditions, reaching the SEN of 99.42±0.48%, PPV of 99.59±0.48%, ACC of 99.02±0.68% and F1 score of 99.50±0.34% even at the lowest noise level.

Figure 6 demonstrated $SNR_{fn}$ acquired by applying different methods with various initial $SNR_{mn}$ levels. The proposed method exhibited obvious $SNR_{fn}$ lift, evidently surpassing those achieved by most of other methods. Occasionally, some methods even resulted in a $SNR_{fn}$ decreasing rather than improvements under poor $SNR_{mn}$ conditions. When the initial $SNR_{mn}$ was relatively high (not less than 3 dB), the CycleGAN method showed performance similar to the proposed method and even achieved greater $SNR_{fn}$ improvements (when the initial $SNR_{mn}$ was 6 dB or 0 dB). However, as the initial $SNR_{mn}$ deteriorated below 0 dB, the proposed method still exhibited a significant advantage over the CycleGAN method.

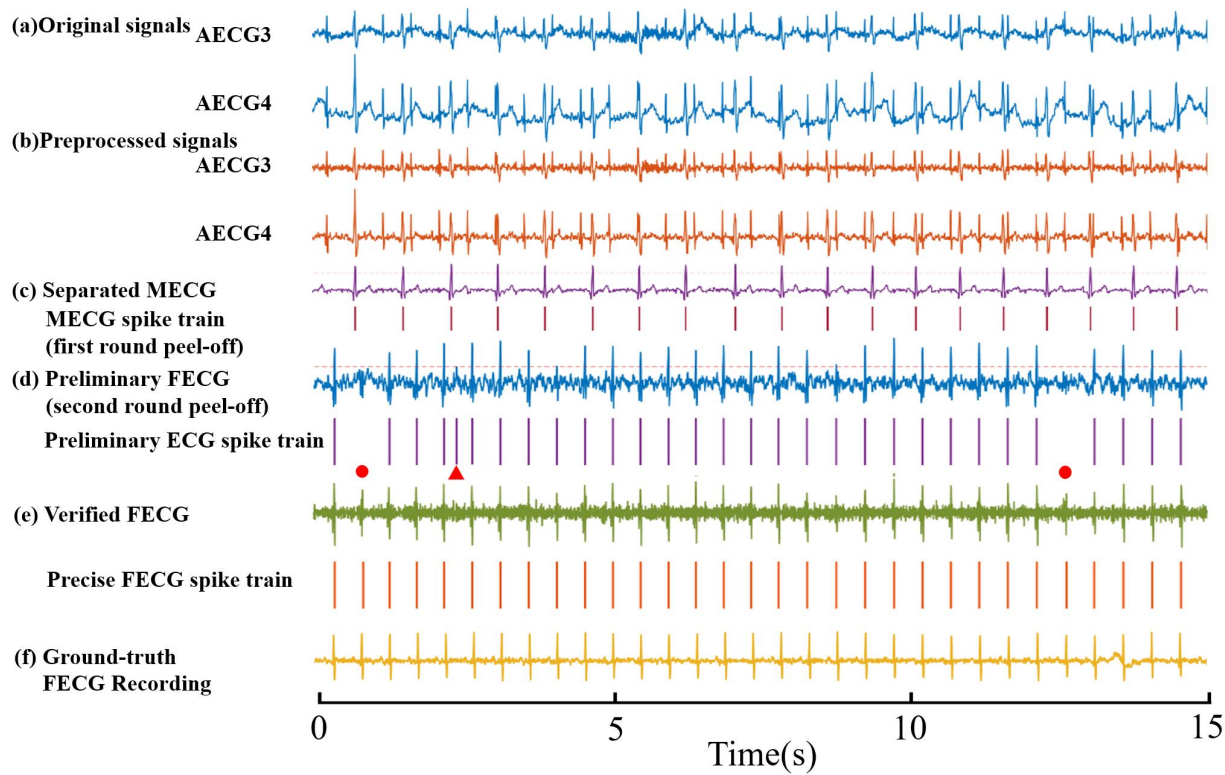

Figure 3. An example of extraction of both MECG and fECG from one representative 15-s AECG data segment in the public database. (a) The original AECG signals in two typical channels. (b) The preprocessed signals in both corresponding channels. (c) The extracted source signal representing MECG and its detected spike train. (d) Another source signal representing preliminarily extracted fECG and its detected spike train with some missing (denoted as a red dot) or erroneous (denoted as a red triangle) spikes. (e) The verified fECG signal and its precise fECG spike train, after applying periodic constrained FastICA with preliminary ECG spike train in (d) used as a reference. (f) The ground-truth fECG recordings for validating the precision of detected fECG signals.

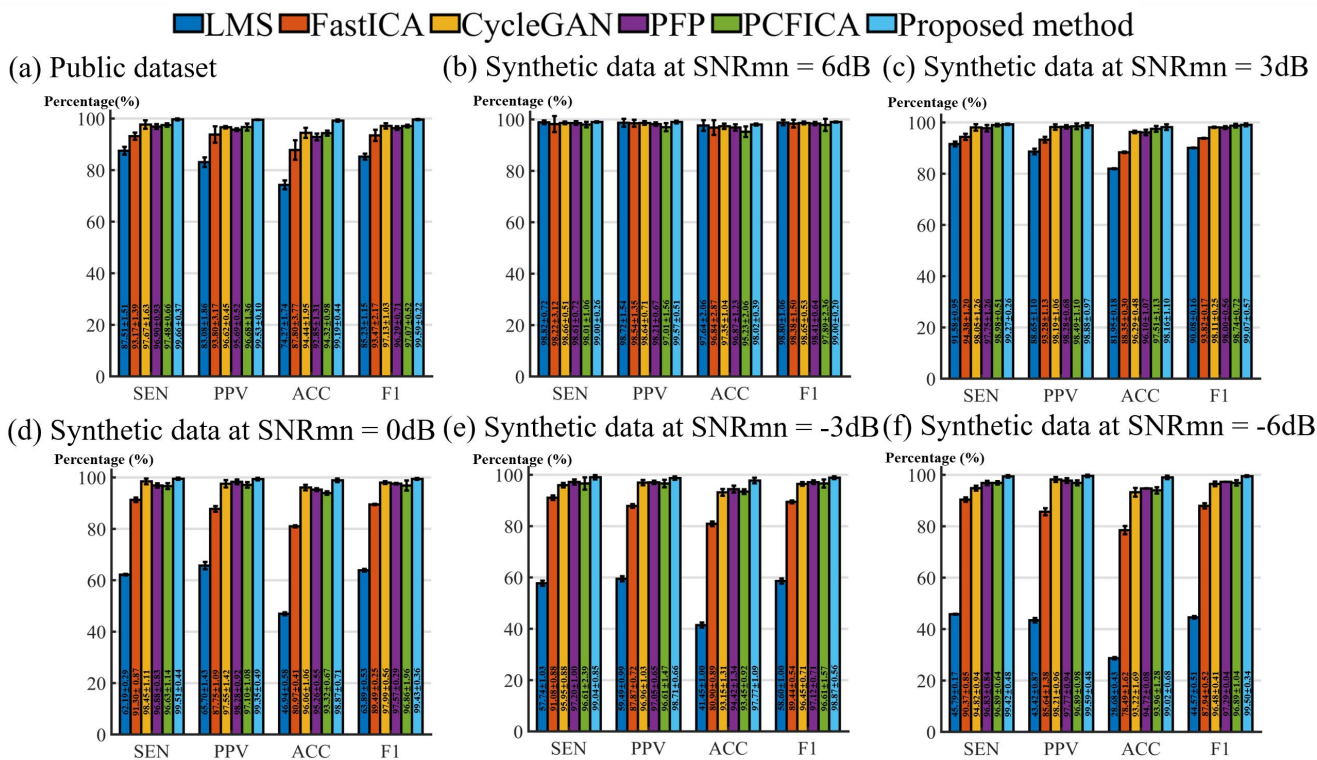

Figure 4. The results of statistical analysis on the processing efficacy using different methods on public datasets and synthetic data. (a) The statistic metrics on public dataset. (b) The statistic metrics on synthetic data with SNRmn of 6 dB. (c) Synthetic data with SNRmn of 3 dB. (d) Synthetic data with SNRmn of 0 dB. (e) Synthetic data with SNRmn of -3 dB. (f) Synthetic data with SNRmn of -6 dB.

To examine the waveform reconstruction performance between SVD and least squares, Figure 7 reports the RMSE values between the reconstructed signal and the pure FECG at different $SNR_{mn}$ levels when using the proposed method (with SVD for waveform reconstruction) and the PCFICA method (with least squares for waveform reconstruction), respectively. The SVD involved in the proposed PPSP method yielded smaller RMSE values compared to the least squares method in PCFICA, across all $SNR_{mn}$ levels with statistical significance

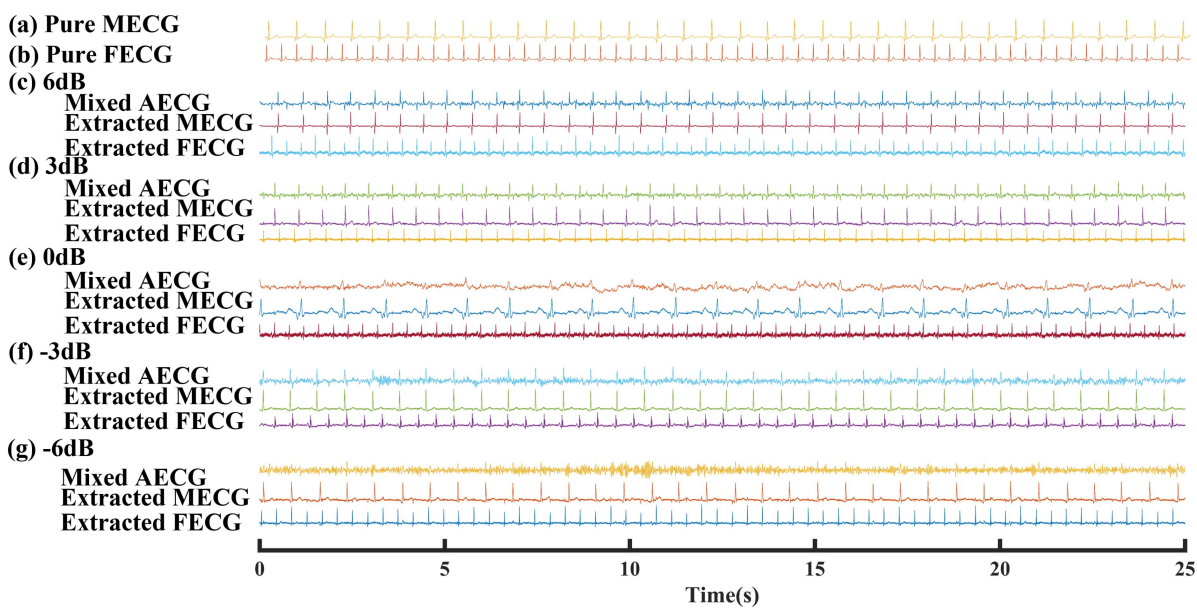

Figure 5 **(a)** Pure MECG. **(b)** Pure fECG. **(c)** AECG with SNRmn of 6dB, corresponded extracted MECG and extracted fECG. **(d)** AECG with SNRmn of 3dB, corresponded extracted MECG and extracted fECG. **(e)** AECG with SNRmn of 0dB, corresponded extracted MECG and extracted fECG. **(f)** AECG with SNRmn of -3dB, corresponded extracted MECG and extracted fECG. **(g)** AECG with SNRmn of -6dB, corresponded extracted MECG and extracted fECG

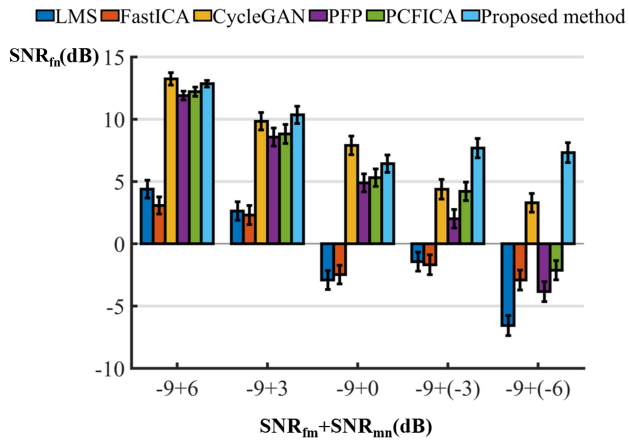

Figure 6. The result of $SNR_{fn}$ of extracted fECG from different noise level acquired using different methods.

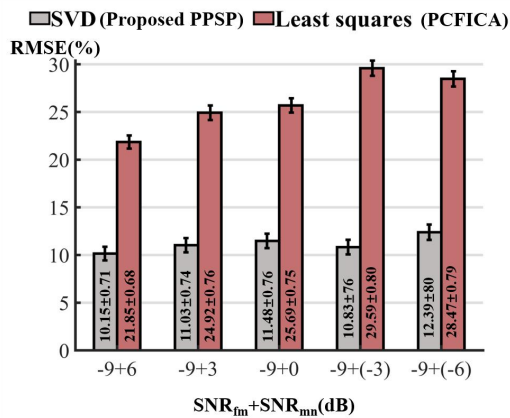

Figure 7. The results of RMSE values between extracted fECG and pure fECG at different noise levels, using PCFICA and proposed PPSP methods.

($p < 0.05$), indicating superior performance in waveform reconstruction.

### C. Results of Clinical Data

Figure 8 shows an example of clinical AECG recordings for both mECG and fECG extraction using the proposed method. It can be observed that the original AECG signal measured under clinical conditions exhibits baseline drift, indicating a significant presence of motion artifacts, and is superimposed with dense and large mECG spike sequences. The fECG waveform is not clearly visible from the typical two AECG channels. The proposed PPSP method can extract smooth and regular mECG waveforms, as well as accurate mECG spike trains. Based on this, it further isolates and extracts clear fECG waveforms and firing sequences. Corresponding to the original AECG sequence, the extracted fECG signal calculates the fetal heart rate, which fluctuates slowly between 120 and 130 beats per minute, with an RMSE of 0.1298 between the FHR curve compared to the actual clinical measurements.

Figure 9 shows the RMSE between FHR obtained from different comparison methods with the FHR curve collected by the clinical ultrasound monitor. It can be seen that the proposed method yields the smallest RMSE for the curve, shows a significant difference compared with other methods.

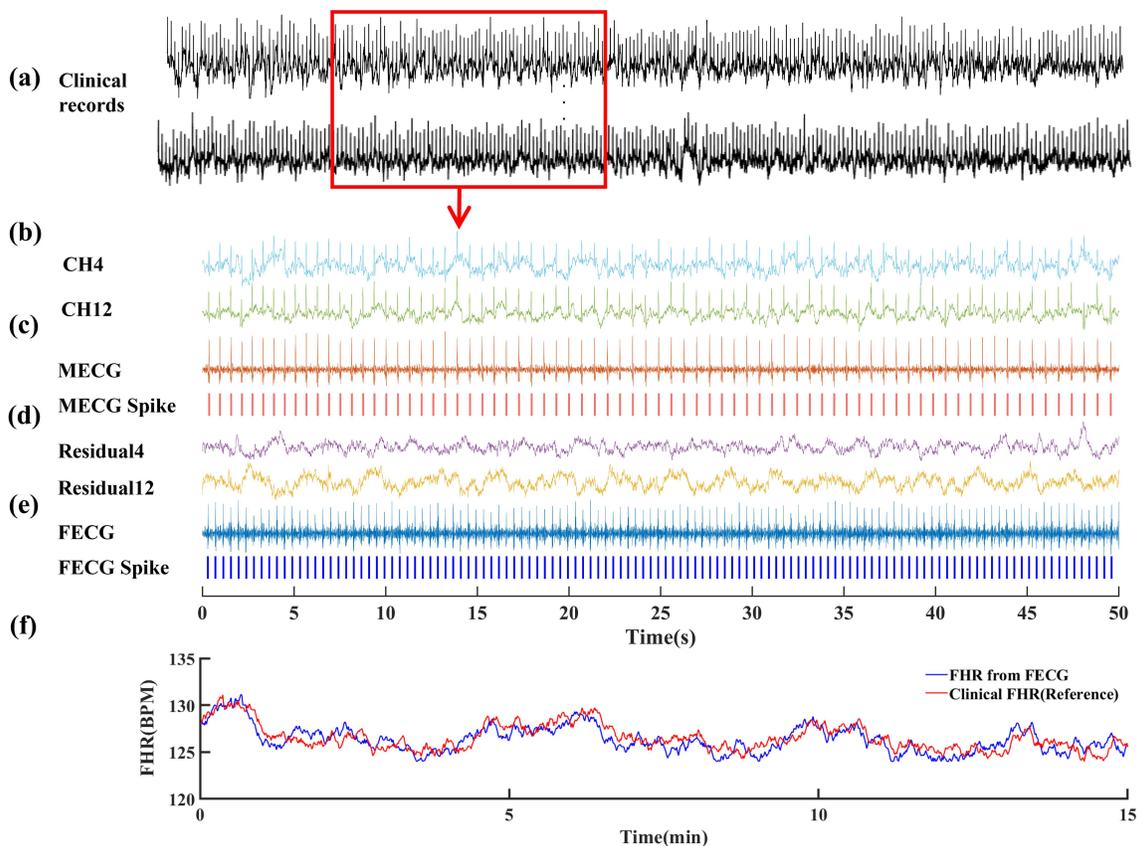

Figure 8. (a) A group of the overall clinical data. (b) Two channels of a 50 second segment from clinical data. (c) Extracted MECG with its spike. (d) Two channels from the 16-channel residual signal obtained through MECG peel-off. (e) Extracted fECG from residual signals. (f) An example of FHR from clinical fECG data, with the reference from clinical ultrasound monitor.

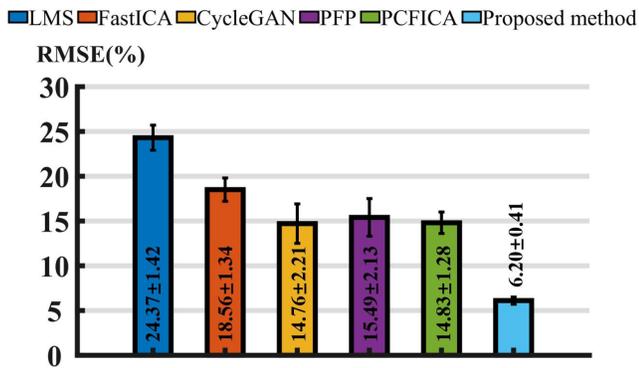

Figure 9. RMSE values between the real FHR and FHR from extracted fECG signals using different methods.

## IV. DISCUSSIONS

The fECG measurement plays a crucial role in fetal activity monitoring. In this study, we present a novel PPSP framework to extract the weak fECG signals from compound signals acquired from high noise environment. The proposed method consists of three primary modules with specific contributions as follows: 1) A two-step FastICA process including a PCFICA module with a periodicity constraint based on the physiological characteristics of ECG, to address the difficulty in accurately extracting the spike train of an ECG source. 2) An SVD-based module for waveform estimation, to solve the problem of reconstructing precise ECG waveforms due to their variations across different spikes and channels. 3) A peel-off strategy, which helps to eliminate the influence of stronger noise sources (like mECG), to address the challenge of extracting dramatically weaker fECG source.

The results on public datasets showed that our proposed method achieved the best performance among all methods in terms of all four metrics (as shown in Figure 4a), demonstrating its effectiveness and strong capability in precisely extracting the fECG spike train. Furthermore, the tests on public databases with clinical benchmark provided a straightforward way to compare our method with previous methods. Table 2 lists a series of previously reported methods and their corresponding metrics. It can be observed that SOTA methods are mainly categorized into two classes: blind source separation (or matrix decomposition) and deep learning, both achieving satisfactory results (all four metrics exceeding 95%). In comparison, our method exhibits evident improvements across all metrics, approximating to almost 100%, which indicates its advantages in the fECG extraction.

It is noteworthy that the public datasets have relatively high signal quality, with the fECG signals clearly visible in the raw mixed AECG signals (shown in Figure 3a). This level of SNR is likely to be consistent with the highest SNR level in our synthetic data (shown in Figure 5c), which indicates that the data collection environment for the public dataset is a relatively ideal scenario. The actual clinical data collection environment, however, is expected to carry more interferences, posing greater challenges for the fECG extraction. This was confirmed by many previous studies. For instance, studies [20] and [44] in Table 2 demonstrated evident performance compromise on private data with respect to that on public datasets. This is also the reason for explaining why the proposed method was tested with synthetic data at lower SNR levels, as shown in Figure 4 (b)- (f). It was unsurprising that all comparative methods and the proposed method achieved good performance at an SNR level of 6dB. However, as the SNR gradually decreased to -6dB, the LMS method exhibited the most noticeable performance degradation. This decline is attributed to its reliance on morphological estimation based on the reference signal and its subsequent subtraction from the original signal, making it unsuitable when noises other than the reference signal become severe. The FastICA method also showed a significant drop in performance, which can be attributed to the fact that it is a basic BSS method without sufficient adaptation to the ECG extraction task. As the first step of the PFP method and the proposed method, it possesses separation capabilities but fails to separate the desired physiological sources. Such separation ability is notably affected as the SNR decreases. On this basis, the performance improvements of the PFP and PCFICA methods validate the effectiveness of adding constraint to the FastICA module. The performance improvement became increasingly obvious as the SNR level decreased, while the PCFICA method with periodicity constraint, which incorporates the physiological characters of ECG signals, performing relatively better than PFP with equality constraint. Additionally, the deep learning method CycleGAN also performed well. In contrast, the proposed method maintained the highest performance metrics even as the SNR declined to a level no more than 0dB, highlighting its advantages under low SNR conditions.

Table 2. Comparative evaluation for the FECG detection performance of different methods on public dataset

| Method | Year | Channel | Database | Recordings | SEN (%) | PPV (%) | ACC (%) | F1 (%) |
|---|---|---|---|---|---|---|---|---|
| TS+PCA[20] | 2014 | 1 | NIFECGA | 14 | 94.7 | 96 | - | 95.4 |
|  |  |  | Private data |  | 89.9 | 88.8 | - | 89.3 |
| Extended Kalman Smoother[44] | 2014 | 3 | NIFECGA | 69 | 97.4 | 97.2 | 96 | 97.3 |
|  |  |  | private data | 24 | 85.8 | 85 | 82.8 | 85.4 |
| FastICA+SVD[42] | 2014 | 4 | NIFECGA | 69 | 99.1 | 98.9 | - | 98.99 |
| ANC+SVD[45] | 2017 | 1 | ADFECG | 2 | 99.37 | 99.49 | 98.90 | 99.45 |
|  |  |  | Private data |  | 98.31 | 98.86 | 97.21 | 98.58 |
| NMF[27] | 2020 | 1 | ADFECG | 5 | 95.3 | 94.6 | - | 94.8 |
| nonlinear estimation[46] | 2019 | 1 | ADFECG | 5 | 93.8 | 98.48 | 92.41 | 96.04 |
|  |  |  | NIFECGA | 15 | 98.63 | 99.52 | 97.77 | 98.85 |
| ST+Shannon Energy[16] | 2022 | 1 | ADFECG | 4 | 96.6 | 96.6 | 100 | 98.27 |
|  |  |  | NIFECGA | 20 | 97.37 | 98.61 | 98.72 | 98.67 |
| CycleGAN[25] | 2021 | 1 | ADFECG | 5 | 99.4 | 99.6 | - | 99.7 |
|  |  |  | NIFECGA | 68 | 96.8 | 97.2 | - | 97.9 |
| DP-LSTM network[10] | 2022 | 1 | ADFECG | 22 | 97.3 | 98.09 | 95.53 | 97.7 |
|  |  |  | NIFECGA | 69 | 94.2 | 96.5 | 91.34 | 95.3 |
| PA$^2$Net[26] | 2022 | 1 | ADFECG | 5 | 99.58 | 99.67 | - | 99.62 |
|  |  |  | NIFECGA | 68 | 98.9 | 98.83 | - | 98.86 |
| CA-KICA[47] | 2023 | 4 | ADFECG | 5 | 98.4 | 97.6 | - | 98.0 |
|  |  |  | NIFECGA | 69 | 99.3 | 99.6 | - | 99.5 |
| **Proposed Method** | **2024** | **4** | **ADFECG** | **5** | **99.71%** | **99.44%** | **99.16%** | **99.58%** |
|  |  |  | **NIFECGA** | **68** | **99.25%** | **99.50%** | **98.77%** | **99.38%** |

Besides, in the results of synthetic data, the $SNR_{fn}$ of the extracted fECG waveforms further measures the performance of fECG waveform extraction (as shown in Figure 6). It can be observed that the proposed method exhibits a significant superior performance under low SNR conditions compared to other methods, with statistical significance. Notably, at high SNR levels, the CycleGAN method even achieved a higher SNR, indicating the advantages of deep learning methods in processing single-channel signal. This advantage is built on extensive data training and complex computations, but it may be affected when processing low SNR signals. The proposed method outperformed CycleGAN under low SNR conditions, confirming its stability and robustness in achieving excellent performance in high-noise environments. This advantage further ensures the practicality of the proposed method in clinical applications.

When dealing with synthetic signals, the proposed method yields fECG with a lower waveform RMSE compared to the PCFICA method, regardless of the SNR conditions (as shown in Figure 7). This indicates the effectiveness of the SVD waveform reconstruction module designed in our PPSP method, which is more suitable for extracting fECG waveforms than the default least square waveform estimation module. This is because it is difficult to satisfy the assumption in the original data model that the sum of noises is white Gaussian noise with zero mean, as the AECG is interfered by various unpredictable noises like EMG noise and fetal movement. When directly estimating the waveform using the ECG spike train, the noise around the spike may be considered as the desired waveform, leading to inaccurate extraction. When using SVD, we applied a trapezoidal window (whose length depended on the mean RR-interval on the whole record) to select and weight the signal around each detected maternal QRS. This operation allows us to avoid artifacts due to abrupt signal truncation [42]. Therefore, the extracted signal has a better degree of reconstruction, leading to a better fECG extraction performance. Extensive testing on both public datasets and synthetic datasets has demonstrated the effectiveness and accuracy of the proposed PPSP method.

The experimental data collected in our clinical practice had a low signal quality, making the fECG not visually apparent in the waveforms, with the occurrence of large mECG spikes in a repetitive pattern. Nevertheless, our PPSP method was able to progressively extract both mECG and fECG, particularly revealing the weak fECG signals and providing a clear sequence of ECG spikes. By calculating the FHR and comparing it with clinical records, the proposed method also achieved the smallest RMSE (as shown in Figure 9), confirming its usability and excellent performance in practical fECG extraction.

The great performance of processing clinical signals also demonstrates the effectiveness of the peel-off strategy in the PPSP method. When the SNR is very low, it is indeed necessary to first separate the large mECG. Even so, the remaining signal after the first separation does not allow for direct observation of the fECG waveform, as shown in Figure 8 d, indicating that the fECG was still hidden beneath baseline noise and artifacts. By executing the two-step FastICA again, we obtained the source signals of the fECG, which further allowed us to derive precise spike trains and waveforms. The peel-off strategy is an important factor contributing to the performance improvement of the proposed method. This work can also be regarded as a new success of the peel-off strategy in the PFP method [30-33] towards fECG extraction task.

Currently, the limitation of our study is that it's an offline algorithm, hence unable to monitor fetal heart in real-time. However, inspired by many online versions of BSS algorithms including online PFP developed by Zhao et al. [30], it is feasible to design an online fECG extraction algorithm for real-time fetal heart monitoring. This remains our future work.

## V. Conclusion

This research reported a PPSP framework for ECG extraction from the AECG signal. By incorporating the PCFICA and peel-off strategy, the proposed framework shows efficiency in fECG extraction from the multi-channel AECG recordings. The framework was validated on public databases, synthetic data and clinical data. Experimental results demonstrated that our method is a promising technique for improving precision of extracting and separating mECG and fECG signals, especially under clinical high-noisy environments. Our study offers a new way of non-invasive measurement and clinical instrumentation for monitoring fetal and maternal heart activities in perinatal medicine.